\documentclass[twocolumn,showpacs,preprintnumbers,amsmath,amssymb]{revtex4}

\usepackage{graphicx}
\usepackage{dcolumn}
\usepackage{bm}

\begin{document}


\title{The 90/10 phenomenon in directed signed social networks}

\author{Long Guo}
 \email{guolong@cug.edu.cn}
\affiliation{%
School of Mathematics and Physics, China University of Geosciences(Wuhan) \\ Lumo road, 430074, Wuhan, China
}%


\date{\today}

\begin{abstract}We empirical study the signs' property in the directed signed social networks of Slashdot and Epinions by using an reshuffled approach. Through calculating the entropy $S_{out}$ and the giant component $G$, we find an interesting 90/10 phenomenon: each individual labels his/her neighbors as friends with $90\%$ or foes with $10\%$ uniformly random from the macroscopic perspective. We also find that the entropy $S_{out}$ is suppressed by the non-randomness of labeling sign. Our present work can prove how do the randomness and the non-randomness coexist in human behavior of labeling signs, qualitatively.
\end{abstract}

\maketitle


{\it Introduction --} Online social networks, such as the EBay, Epinions and Slashdot, provide the convenient platforms to manage social interactions intrinsically involving both positive and negative relationships. Those social networks can be represented in terms of signed social networks\cite{facchetti2012, esmailian2014, ciotti2015}, where a sign of link is defined as "+" or "-" depending on whether it expresses a positive or negative attitude from the generator of the link to the recipient\cite{leskovec20102}. For example, users can tag directed relations to others indicating trust or distrust in the trust network of Epinions, and users can designate others as "friends" or "foes" in the social network of the technology blog Slashdot\cite{leskovec20101}.

The fundamental and interesting issue is the signs' organization in real signed social networks. There exists two main social-psychological theories of triad types: the status theory\cite{leskovec20102} and the social balance theory\cite{facchetti2011}. The status theory was developed by Leskovec et al.\cite{leskovec20102} and applied in directed signed networks. In the status theory, a positive edge $(u, v)$ means that $u$ regards $v$ as having higher status than himself/herself while a negative edge $(u, v)$ means that $u$ regards $v$ as having lower status than himself/herself. The social balance theory was first articulated by Heider, based on the adages that "the friend of my friend is also my friend", "the enemy of my friend is my enemy" and so on\cite{marvel2009,lerner2016}. The social balance theory has recently attracted more attention from sociologists and physicists, and is applied in the core questions of sign prediction\cite{leskovec20102} and community detection\cite{chen2016}. Both of those theories were considered from the local structure property in signed networks, for instance, triad types in the social balance theory and two-tuples types in the status theory. However, the interplay between positive and negative signs is seldom considered from the isolated individual's viewpoint according to the subjective initiative of human. In other words, how does individual label his/her neighbors as friends or enemies from the statistical mechanics viewpoint in society, especially in online virtual society?

In this paper, we would like to provide a remarkably simple rigorous proof for the randomness of labeling signs in directed signed social networks of Epinions and Slashdot by using the percolation theory and the information entropy. As we know, percolation is one of the best-studied processes in statistical physics\cite{karrer2014} and serves as a conceptual framework to treat more factual problems on networks\cite{serrano2006}, such as the robustness and collective behavior\cite{serrano2011}. And information entropy describes the uncertainty associated with a given probability distribution. The application of entropy concept in complex networks is widely and deeply \cite{bianconi2008, anand2011, zhao2011, ye2014, anand2014}. However, the application of entropy and percolation in signed network is presently limited and challenged. Our motivation in present work is to investigate the signs' labeling property by using the two theories.

{\it Description of signed social networks and our reshuffled approach --} To begin with, we consider a directed signed social network $G=(V, L, A)$, where $V$ is the set of vertices, $L$ is the set of directed links, $A=\{A_{uv}\}$ is the signed adjacency matrix where $A_{uv}\neq 0$ if and only if $(u,v)\in L$, and $A_{uv}$ is the sign of link $(u,v)$. A positive sign $A_{uv}(=+1)$ represents that $u$ tags $v$ as a friend or $u$ trusts $v$, while a negative sign $A_{uv}(=-1)$ reflects that $u$ tags $v$ as a foe or $u$ distrusts $v$. Each directed signed social network has its own macroscopic parameters $(N_{v}, N_{l}, q_{+}^{0})$: the number of vertices $N_{v}$, the number of directed links $N_{l}$ and the ratio of the positive links $q_{+}^{0}$. The trust network of Epinions obtained in August 12, 2003 and the social network of Slashdot obtained in November 6, 2008 \cite{leskovec20101, leskovec20102} will be empirically analyzed in our present work. The trust network of Epinions with the macroscopic parameters $(131 828, 841 372, 85.3\%)$ is a product review Website with a very active user community, where users can tag their trust or distrust of the reviews of others, and the social network of the Slashdot with $(77 357, 516 575, 76.7\%)$ is a technology-related network website, where a signed link means that one user likes or dislikes the comments of another user.

A given signed social network with two constant macroscopic parameters $N_{v}$ and $N_{l}$ can be regarded as an isolated system, which obeys the ergodic hypothesis from the viewpoint of statistic mechanics. According to the ensemble theory in statistical mechanics, each signed social network has $2^{N_{l}}$ possible microstates, which is related to the signs' organization. Different signs' organization describes different microstate. There must exist one specific microstate which has the same signs' organization of the real signed social network. Our main goal here is to analyze how do signs organize in a given real signed social network through comparing entropies $S_{out}$s and the giant components $G$s of all possible signs' organization. In order to do this, we introduce a reshuffled approach to reconstruct configurations of the given signed social network with constant macroscopic parameters $N_{v}$ and $N_{l}$. In our reshuffled approach, each link is chosen as a reshuffled link with probability $p_{s}$, and then each reshuffled link is reset its sign as the positive one (sign "+") with probability $p_{+}$ or the negative one (sign "-") with probability $1-p_{+}$. The reshuffled signed network with pair parameters $p_{s}$ and $p_{+}$ is fixed when the reshuffled process finished, and so the ratio of the positive sign can be written as
\begin{equation}
\label{eq.1}
 q_{+}(p_{+}, p_{s})=q_{+}^{0}+(p_{+}-q_{+}^{0})p_{s}.
 \end{equation}
 The reshuffled signed network is reduced to the real one when $p_{s}=0$, while all signs will be reshuffled thorough randomly when $p_{s}=1$.

{\it Results from information entropy -- } Analogously to an Ising model, a positive link is mapped as one with spin "$\uparrow$" while a negative link mapped as one with spin "$\downarrow$". The presence of negative links introduce disorder (or frustration) in signed social network\cite{facchetti2012}, where sign of link also describes the social property, such as friend and enemy, of the corresponding connection between individuals. Each vertex has his/her social status according to signs of his/her connections\cite{ball2013}. The social status $p_{i(out)}^{+}$ of individual $i$ is defined as the proportion of the positive links directing to his/her local neighbors due to the subjective initiative of person. The difference among individuals' social status, which is described as the distribution of individuals' social status, can reflect the disorder of signed social network. It is obvious that $0\leq p_{i(out)}^{+}\leq 1$ and the distribution of $p_{i(out)}^{+}$ with the bin width $\tau$ is defined as

\begin{equation}
\label{eq.2}
\pi_{j}^{out}=\frac{\sum\limits_{i=1}^{N_{v}}\delta(j\tau\leq p_{i(out)}^{+}\leq(j+1)\tau)}{N_{v}}
\end{equation}
where $\pi_{j}^{out}$ is the probability that each vertex $i$ with $p_{i(out)}^{+}$ falls in the bin of $(j\tau\leq p_{i(out)}^{+}\leq(j+1)\tau)$, and $\delta(x)$ is the step function. $\delta(x)=1$ when the condition $x$ is true, while $\delta(x)=0$ otherwise. When the reshuffled probability $p_{s}$ is given(i.e., the microstate of the system is given), the probability distribution $\{\pi_{j}^{out}|j=0, 1, 2, ..., (\lfloor\frac{1}{\tau}\rfloor-1)\}$ is fixed.

The entropy of the probability distribution $\{\pi_{j}^{out}|j=0, 1, 2, ..., (\lfloor\frac{1}{\tau}\rfloor-1)\}$ is given by:
\begin{equation}
\label{eq.3}
S_{out}(p_{s}, p_{+})=-\sum_{j=0}^{\lfloor\frac{1}{\tau}\rfloor-1}\pi_{j}^{out}log\pi_{j}^{out}
\end{equation}

 Eq.(\ref{eq.3}) describes the disorder degree of vertices' social status and is called the entropy of social status for out-links. Note that the bin width $\tau$ only changes the value of $S_{out}$ but does not change the evolution tendency of $S_{out}(p_{+})$, which is not shown here. Hence, as a concrete example, we choose the bin width $\tau = 0.05$.

\begin{figure}
\resizebox{0.53\textwidth}{!}{%
  \includegraphics{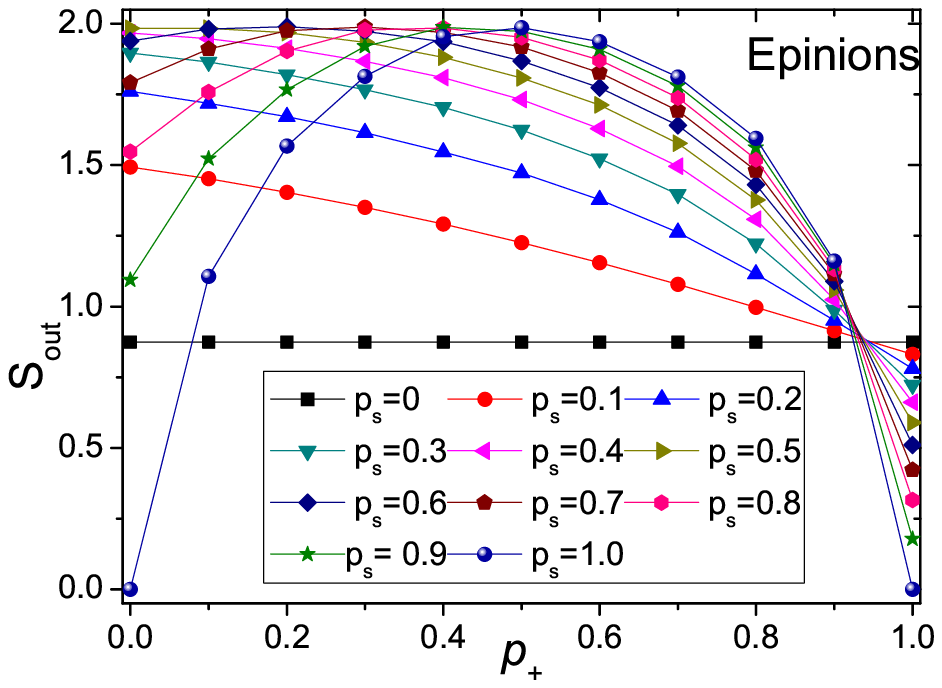}
  \includegraphics{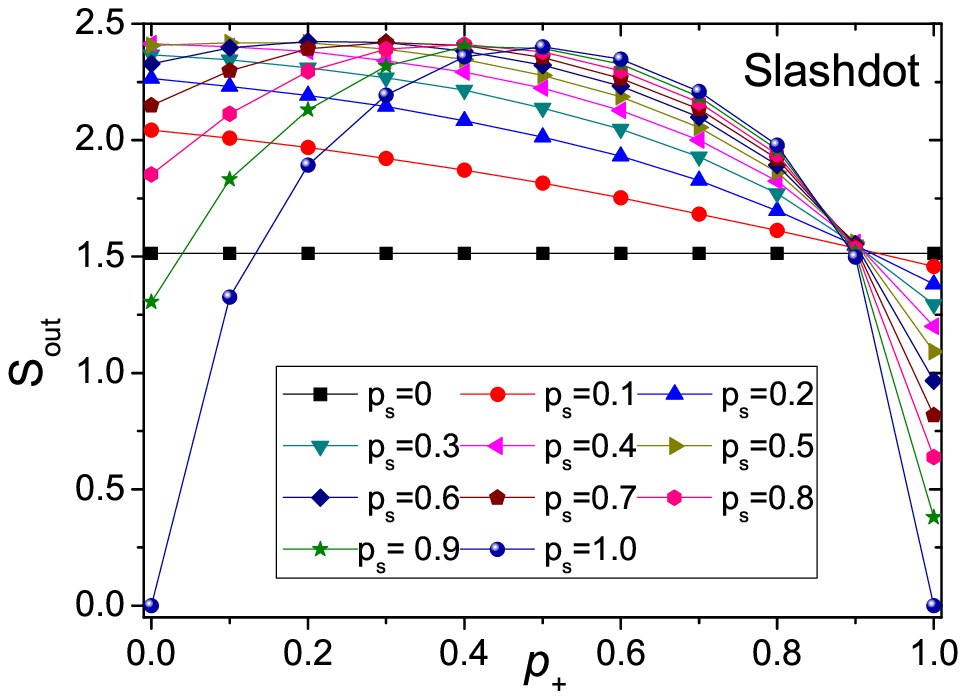}
}
\caption{(Color online)All entropies $S_{out}$ of reshuffled signed networks of Epinions (left picture) and Slashdot (right picture) are plots in the same picture respectively. We focus on the evolution of $S_{out}$ as a function of the resetting probability $p_{+}$ under different reshuffled probabilities $p_{s}$s in detail.}
\label{entropy}
\end{figure}

We study the evolution of $S_{out}$ as a function of $p_{+}$ under each reshuffled probability $p_{s}$ in Fig.\ref{entropy}. $S_{out}$ decreases as $p_{+}$ increases when $p_{s}<0.5$, while $S_{out}$ first increases to its maximum value at $p_{+}=0.5$ and then decreases as  $p_{+}$ increases when $p_{s}>0.5$. All the evolution of those curves share that the real signed social network has the minimum entropy among all the reshuffled signed networks, which is the fingerprint of the homophily property in society. The interesting result is that all curves intersect at the corresponding entropy value of real signed social network at $p_{+}\simeq 0.9$. Namely, all the reshuffled signed network has the same entropy as the real signed network at $p_{+}\simeq 0.9$, which is independent of  $p_{s}$. That is to say, the competition of positive sign and negative one can be described as the 90/10 phenomenon: each individual has about 90\% friends and about 10\% foes among his/her local neighbors.

\begin{figure}
\resizebox{0.53\textwidth}{!}{%
  \includegraphics{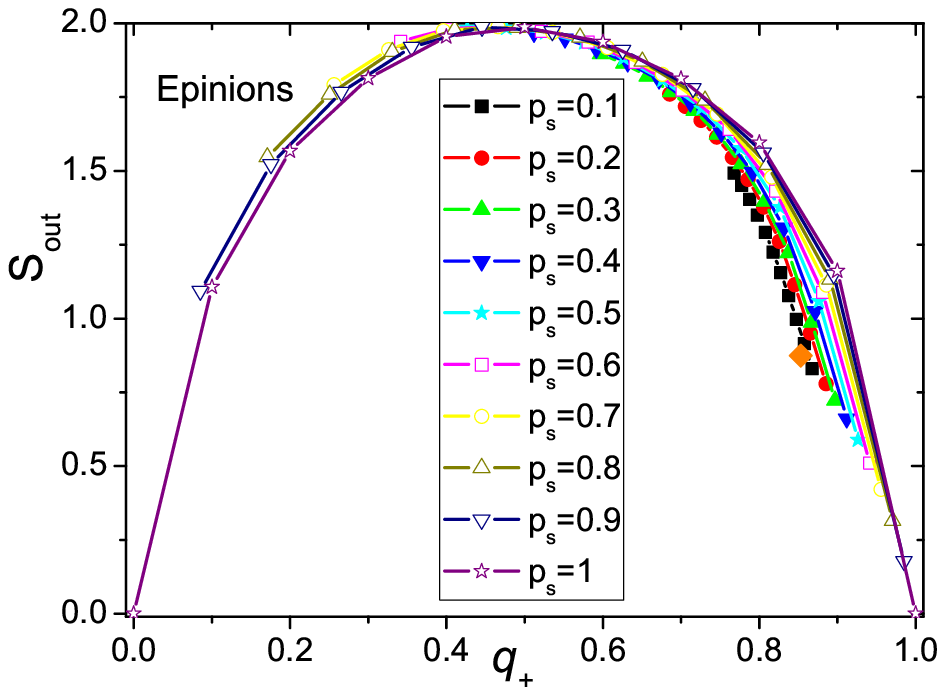}
  \includegraphics{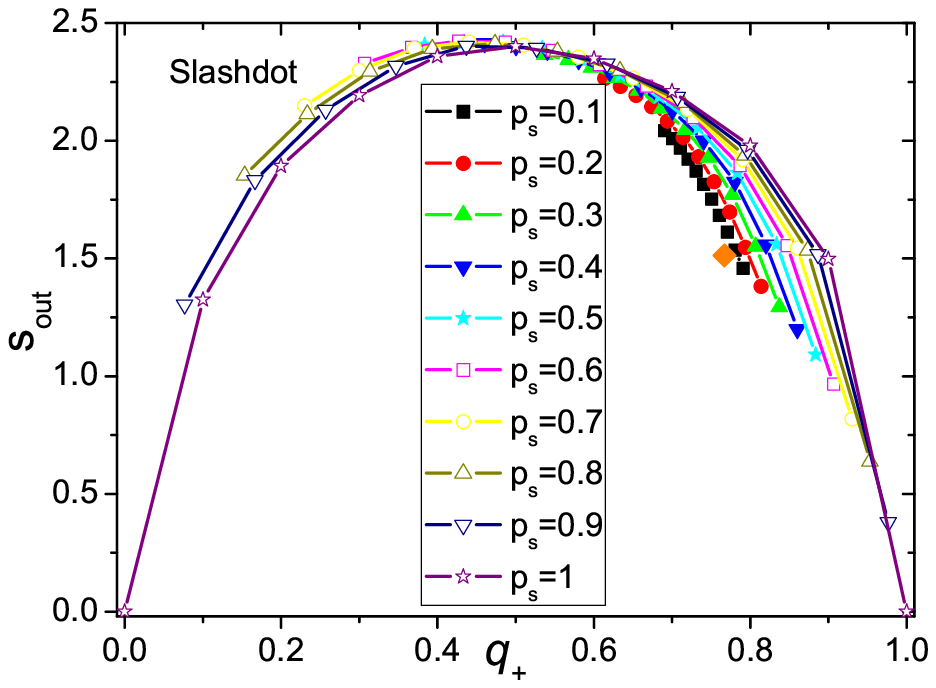}
}
\caption{(Color online)The entropy $S(out)$ versus the ratio of positive signs $q_{+}$ in each reshuffled signed network with different $p_{s}$ is plotted. The orange diamonds are the entropies of the two real signed social networks of Epinions and Slashdot respectively.}
\label{entropyq}
\end{figure}

Next, the 90/10 phenomenon also shows the randomness of labeling sign in society from macroscopic viewpoint. In order to prove the randomness property, we analyze the relationship of the entropy $S_{out}$ and the ratio of positive signs $q_{+}$ (see Eq.(\ref{eq.1}) in reshuffled signed social networks. All signs are reshuffled randomly and the ratio of positive signs is $p_{+}$ in the reshuffled signed network with $p_{s}=1$. Then, we plot the figure of the evolution of $S_{out}$ as a function of $q_{+}$ in Fig.\ref{entropyq}. We find that $S_{out}$ of the reshuffled network with $p_{s}=1$ evolves as a bell-shaped function of $q_{+}$, which is an obvious signal of the randomness of labeling signs. The surprising result is that $S_{out}$ is suppressed in the range of $0.5<q_{+}<1$ when $p_{s}\neq 1$. And signs in reshuffled network with $p_{s}\neq 1$ can be divided into two parts: the reset signs through reshuffled process and the original signs left in real signed network. The smaller $p_{s}$makes the value of $S_{out}$ smaller for each fixed $p_{+}$. $S_{out}$ of the real signed social network is the smallest among those corresponding reshuffled signed networks with the same $q_{+}=q_{+}^{0}$, see the position of the orange diamonds in Fig.\ref{entropyq}. The suppressed phenomenon is the fingerprint of the non-randomness during labeling signs in society and we do not analyze the details of non-randomness in our present work.

{\it Results from percolation theory -- } We can also analyze the randomness of signs' labeling in directed signed social network in analogy to the bond percolation, because the positive sign or negative sign of link can be mapped as the occupied or unoccupied state of link. In the bond percolation process, each link is occupied uniformly at random with probability $p$ or unoccupied with probability $1-p$, just as each link is labeled as positive sign with probability $q_{+}$ or negative sign with probability ($1-q_{+}$) in reshuffled signed social network with $p_{s}=1$. The primary entities of interest are the percolation clusters\cite{karrer2014}, sets of nodes connected by positive links. We here calculate the biggest cluster, also called the giant component $G$\cite{breskin2006, cho2009}, in each reshuffled signed network with the pair parameters ($p_{s}, p_{+}$). We adopt our reshuffled approach to build a reshuffled signed network and then calculate $G$ of each given reshuffled network.

\begin{figure}
\resizebox{0.53\textwidth}{!}{%
  \includegraphics{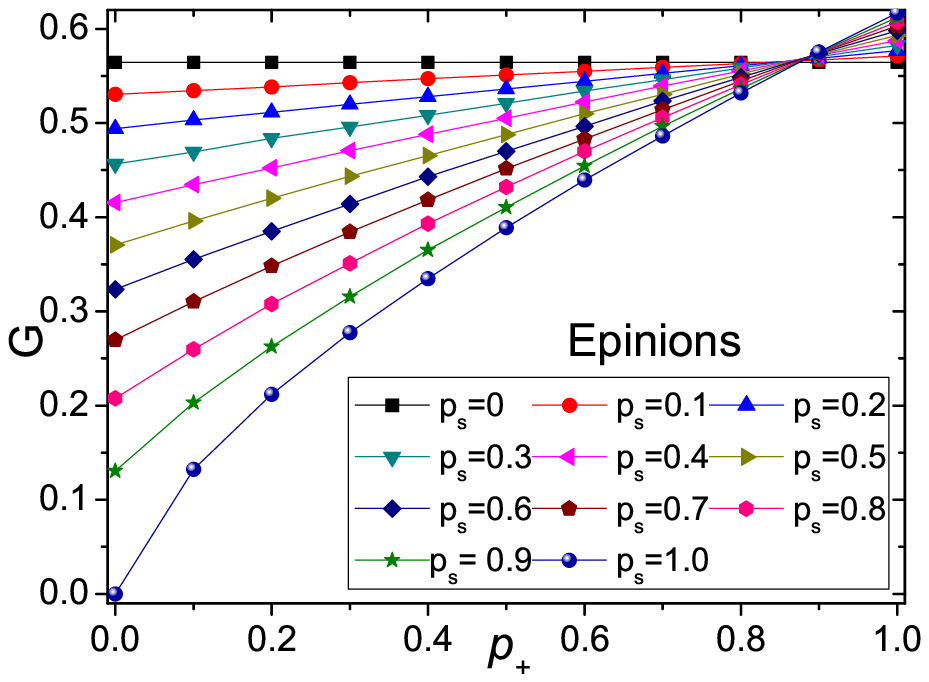}
  \includegraphics{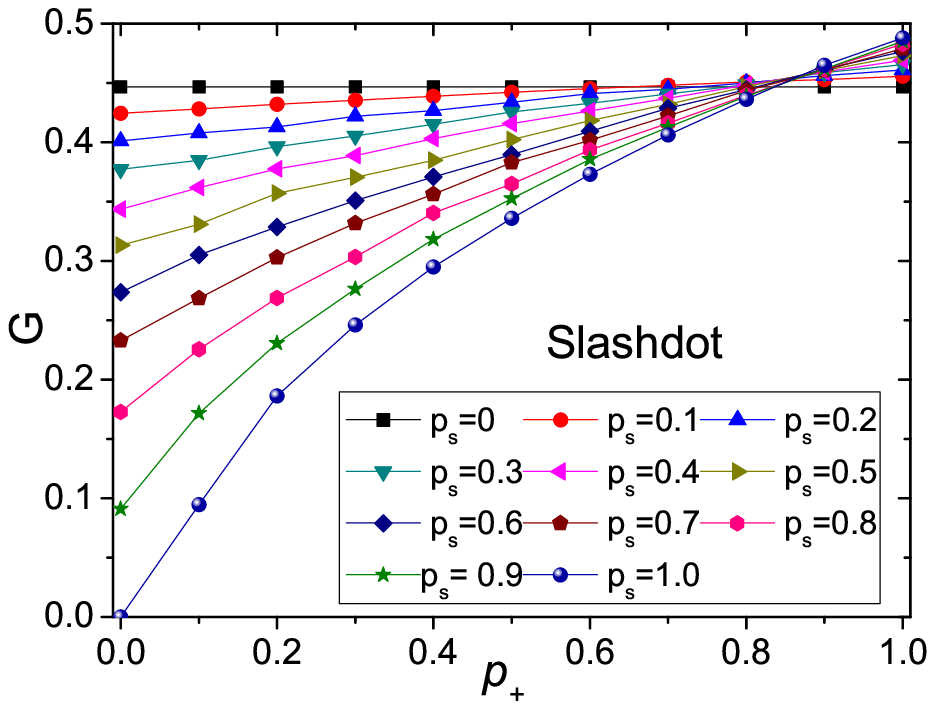}
}
\caption{(Color online)The giant component $G$ versus the resetting probability $p_{+}$ in reshuffled signed networks of Epinions (left picture) and Slashdot (right picture) under different $p_{s}$. }
\label{elarclu}
\end{figure}

In Fig.\ref{elarclu}, all $G$s' of reshuffled signed networks of Epinions and Slashdot with different parameters $p_{s}$ and $p_{+}$ are plotted in the same picture respectively. We find that all $G$s of those reshuffled signed networks are smaller than that of the real signed social network when $p_{+}<0.9$, which means that human activity of labeling friends or foes enhances the giant component formation in real signed social networks and proves the herd effect in society. Comparing $G$s of reshuffled signed social networks with that of the special reshuffled signed social network of $p_{s}=1$, smaller $p_{s}$ leads to larger $G$, which reflects the role of non-randomness of labeling signs in real signed social networks. The most surprising issue is that we also find the signal of the 90/10 phenomenon, since all $G$s are the same as the value of $G$ in real signed social network when $p_{+} \simeq 0.9$.

\begin{figure}
\resizebox{0.53\textwidth}{!}{%
  \includegraphics{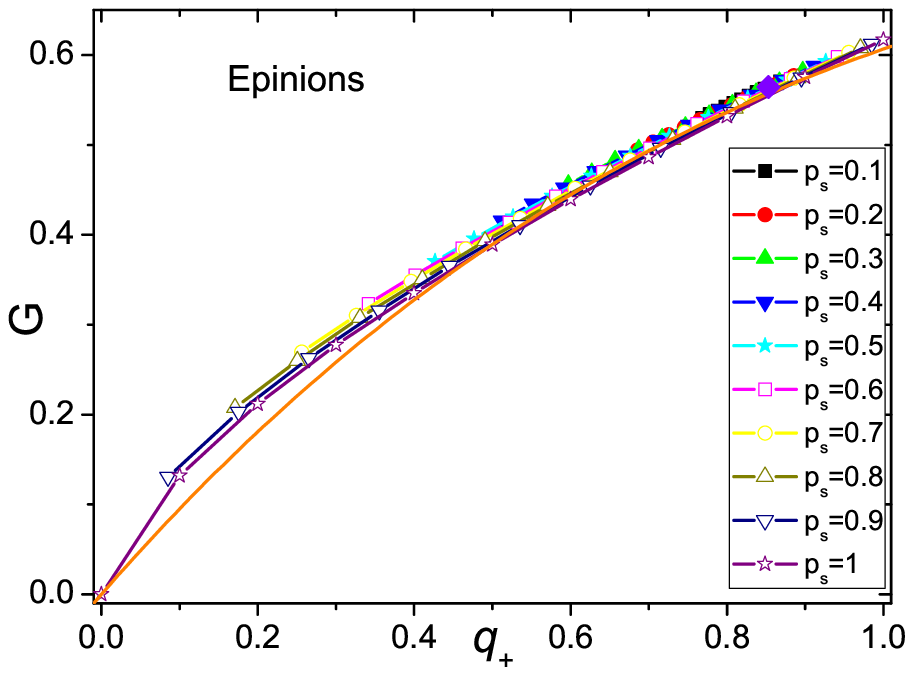}
  \includegraphics{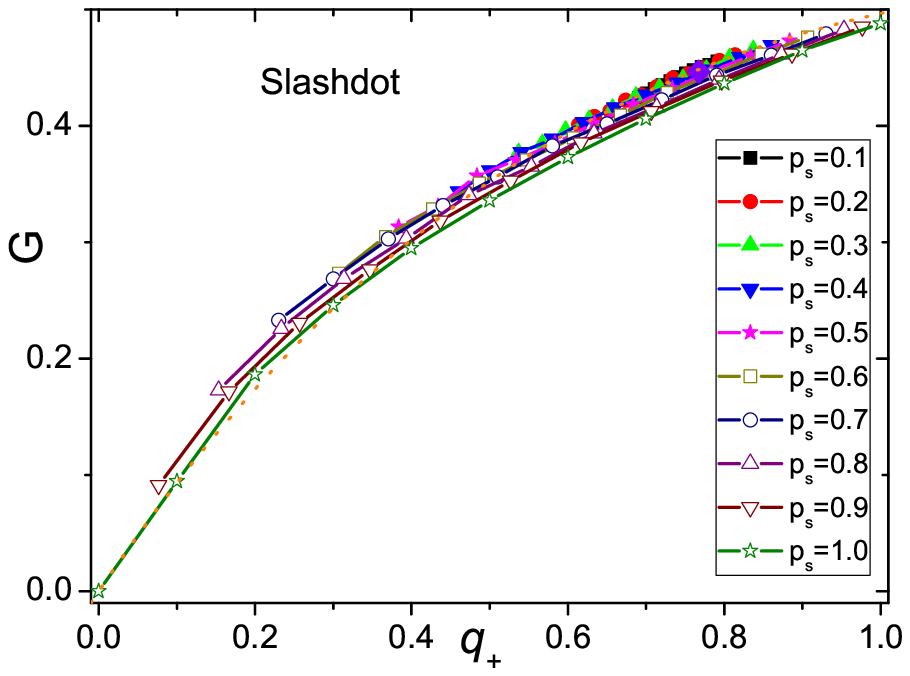}
}
\caption{(Color online)The giant component $G$ versus the ratio of positive signs $q_{+}$ in reshuffled signed networks of Epinions (left picture) and Slashdot (right picture) under different $p_{s}$.The orange solid curve is the fitting one of $G=q_{+}e^{-o.5q_{+}}$ in the reshuffled signed networks of Epinions with $p_{s}=1$, and the orange dash curve is the fitting one of $G=q_{+}e^{-o.7q_{+}}$ in the reshuffled signed networks of Slashdot with $p_{s}=1$. The two violet diamonds here are the giant components of the two real signed social networks of Epinions and Slashdot respectively.}
\label{slarclu}
\end{figure}

 Furthermore, all those reshuffled signed networks have the same topology structure if we do not consider signs of links. In the special case of $p_{s}=1$, each link is labeled as positive sign uniformly random with probability $q_{+}$($=p_{+}$) or negative sign with probability ($1-q_{+}$). In Fig.\ref{slarclu}, we plot the evolution of $G$ as a function of $q_{+}$ in the special case of $p_{s}=1$ and find that the fitting function $G=q_{+}e^{\alpha q_{+}}$, where $\alpha=0.5$ for the Epinions network and $\alpha=0.7$ for the Slashdot network. For other $p_{s}$($\neq 1$)s, all curves of $G$, also including $G$s of the two real directed signed social networks of Epinions and Slashdot (see the violet diamonds in Fig.\ref{slarclu}), overlap with each other shows the randomness of labeling signs in society since the Giant component has nothing to do with the reshuffled probability $p_{s}$.  Our result of the randomness of labeling signs provides a strong proof for the feasible of studying society using incomplete information.

{\it Conclusion and discussion -- } In summary, we have empirical analyzed the property of labeling signs in the two directed signed social networks of Epinions and Slashdot by calculating the entropy $S_{out}$ and the giant component $G$. In order to show the plausible signs' organization in real signed social networks considering the individual's aspect, we build many configurations of those two real signed networks using the reshuffled approach. In reshuffled approach, each link is chosen to be the reshuffled one with the reshuffled probability $p_{s}$, then each reshuffled link is reset its sign as positive one with probability $p_{+}$ or negative one with probability ($1-p_{+}$). For each configuration with pair probabilities ($p_{s}, p_{+}$), $S_{out}$ and $G$ are calculated. We find that each individual label each one of his neighbors as friend uniformly random with probability 90\% or enemy with probability 10\% from global aspect. In addition, the suppressed phenomenon of $S_{out}$ shows the role of non-randomness of human behavior in society. It will be interesting to note that our results of the randomness and the 90/10 phenomenon maybe prove the scientific feasible of researching human behavior using local or incomplete information of society. Our present work provide an interesting tools and perspective to analyze links' property and function in social network.

\bibliography{apssamp}

\end{document}